# Deep learning reconstruction of ultrashort pulses from 2D spatial intensity patterns recorded by an all-in-line system in a single-shot

RON ZIV[1], ALEX DIKOPOLTSEV[2], TOM ZAHAVY[1], ITTAI RUBINSTEIN[2], PAVEL SIDORENKO[2], OREN COHEN[2] AND MORDECHAI SEGEV[2]

[1] Department of EE, Technion, Haifa 32000, Israel
[2] Department of Physics and Solid State Institute, Technion, Haifa 32000, Israel
*msegev@technion.ac.il

**Abstract:** We propose a simple all-in-line single-shot scheme for diagnostics of ultrashort laser pulses, consisting of a multi-mode fiber, a nonlinear crystal and a CCD camera. The system records a 2D spatial intensity pattern, from which the pulse shape (amplitude and phase) are recovered, through a fast Deep Learning algorithm. We explore this scheme in simulations and demonstrate the recovery of ultrashort pulses, robustness to noise in measurements and to inaccuracies in the parameters of the system components. Our technique mitigates the need for commonly used iterative optimization reconstruction methods, which are usually slow and hampered by the presence of noise. These features make our concept system advantageous for real time probing of ultrafast processes and noisy conditions. Moreover, this work exemplifies that using deep learning we can unlock new types of systems for pulse recovery.

## 1. Introduction

Ultrashort femtosecond-scale laser pulses are a key ingredient in time-resolved investigations of ultrafast phenomena [1–3], such as chemical reactions, electron dynamics in atoms and molecules etc., hence their complete characterization (amplitude and phase) is of great importance. However, sensor technology does not yet have short enough response time to recover ultrashort pulses directly. Consequently, ultrashort pulses are currently recovered indirectly, often through algorithmic methods. A widespread method is the Frequency-resolved optical gating (FROG [4]) which is based on gating a pulse with a time shifted replica of itself inside a nonlinear medium, and measuring the power spectrum of the nonlinear signal as function of the delay between the pulses (the 2D FROG trace). A more recent method that gains popularity, known as d-scan (dispersion scan [5]), is based on varying the dispersion experienced by the probe pulse before the pulse passes through the nonlinear crystal. D-Scan relies on forming a 2D trace by stacking the nonlinear spectra at the different degrees of dispersion. The reconstruction algorithms used in both FROG and d-scan to recover the probed pulse from the recorded spectrograms are iterative phase-retrieval algorithms. Recently, with the progress in machine learning [6,7], deep-learning-based reconstruction of ultrashort pulses was demonstrated in both FROG [8], d-scan [9] and other techniques [10], displaying considerable improvements in terms of speed of reconstruction and noise robustness, due to the intrinsic ability of deep learning to filter out noise [11]. However, in the field of diagnostics of ultrashort laser pulses, deep learning techniques were so far employed only for improving the performance of already existing schemes, but never as the pulse recovery method associated with a completely new scheme.

Conventional FROG and d-scan devices work in the multi-shot regime, which requires trains of (almost) identical pulses. However, in some experiments [12,13], the probed pulse is not part of a train of identical pulses, hence it is often desired to characterize a pulse using a single-shot characterization method. Indeed, single-shot variants of FROG (termed GRENOUILE [14]) and of d-scan [15] were developed. In GRENOUILE, a Fresnel bi-prism splits an incoming pulse beam into two non-collinear stripe beams, where the delay between the pulses is mapped to a spatial axis. The beams overlap in a thick nonlinear crystal, acting also as a spectrometer by utilizing the narrow bandwidth frequency-angle phase-matching dependence. In this scheme, the group velocity mismatch and crystal dispersion result in a tradeoff between the spectral bandwidth and temporal duration of the pulse, limiting the time-bandwidth product of the pulses such a device can recover. In single-shot d-scan, different transverse parts of the beam experience different degrees of dispersion by using a prism and imaging system. The number and accuracy of sampled dispersion points is limited by the resolution of the imaging system. Another single-shot method is Spectral Phase Interferometry for Direct Electric—Field Reconstruction (SPIDER [16]), which is based on spectral interferometry between the probed pulse and a frequency shifted replica of itself. The probed pulse is calculated directly from the measured 1D inerferogram. However, SPIDER devices are generally more complicated and expensive than GRENOUILE and single-shot d-scan. Also, generally single-shot devices are inherently highly sensitive to noise (as there is no averaging by multiple pulses and the power of the incoming signal is always limited), hence noise robustness of the reconstruction algorithm is especially critical in single-shot pulse recovery systems.

Here, we propose and demonstrate in simulations, a simple single-shot pulse characterization system, based on all-in-line propagation of the pulsed beam using off-the-shelf fibers and a $\chi^{(2)}$ nonlinear crystal. The output of the crystal is imaged onto a CCD that records a 2D speckle intensity pattern. We show that by using deep learning techniques, we can successfully reconstruct the pulse from the recorded data, even at low SNR, a previously unattainable feat for these types of measurement systems. We prove that this system can provide a practicable single shot measurement apparatus for diagnosis of ultrashort pulses, and validate the robustness of the reconstruction to physical variations in the system. Last but not least, the multi-mode fiber in our scheme can be replaced by other components that mix the beam's spectral and spatial degrees of freedoms (e.g. a thick diffuser), which would offer greater resolution and dynamic range.

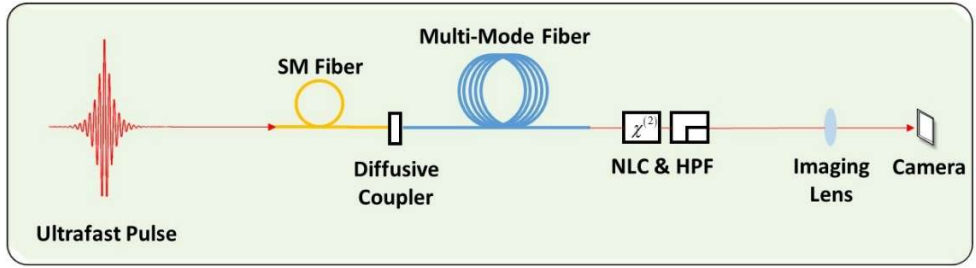

Fig. 1 Proposed system for ultrafast pulse measurement. The ultrashort pulse is passed through a single mode (SM) fiber, and subsequently through a diffuser that couples to multiple modes of a multimode fiber. The resultant spatio-temporal pulse exiting the multimode fiber is passed through a nonlinear crystal (NLC) and then through a high-pass frequency filter (HPF) that passes only the sum-frequencies. The ensuring interference pattern is imaged onto a camera.

## 2. Proposed system

The system is based on spatiotemporal coupling and a nonlinear measurement (see Fig.1). The first element of the system is a single-mode fiber. When a pulse enters the single-mode fiber, its spatial profile is coupled (projected) to the spatial mode of the single-mode fiber, while retaining its spectrum and spectral phase. We assume that the field propagates linearly in the single-mode fiber, experiencing only dispersion. Next, for the purpose of creating spatiotemporal coupling, we use a diffusive coupler and a multi-mode fiber. The diffusive coupler, projects the field from the single-mode fiber onto the different modes of the multi-mode fiber, exciting each mode with a different amplitude and phase. In the multi-mode fiber, each mode has a different spatial profile and a different rate of phase accumulation according to the modal dispersion and the wavelength. Next, we introduce a $\chi^{(2)}$ nonlinear medium with a large enough numerical aperture such that the highly multimode field emerging from the fiber is coupled to the crystal without being truncated. The electric field emerging from the crystal is composed of the outcome of all the $\chi^{(2)}$ processes occurring among the frequency components of the pulse: sum frequency generation, difference frequency and rectification (for simplicity, we neglect cascaded effects where $\chi^{(2)}$ operates twice or more). The emerging field is then passed through an optical high pass filter, keeping only the sum frequency elements (SFG). This complex high-frequency field is then imaged onto a camera that measures the time-averaged interference pattern created by the sum-frequency field of the pulse. The nonlinearly of this process, combined with the spatiotemporal mixing due to the diffusive coupler and the propagation through the multi-mode fiber, creates an interference pattern that depends on the amplitude and spectral phase of the pulse. As described below, this interference pattern allows for deciphering the amplitude and phase of the incoming pulse.

We choose this system due to its simplicity and small size, and because it offers an easily controllable method of interference between different frequencies, by simply changing the propagation length in the dispersive medium (multi-mode fiber). We note that such a system, but without the nonlinearity, was used for high resolution low loss spectrometry [21]. However, that system measures only the power spectrum of the pulse while the spectral phase is lost, and without it complete pulse reconstruction is not possible.

Next, we analyze our proposed system theoretically and simulate examples. To that end, we calculate the spatiotemporal coupling of the pulse emerging from the single-mode fiber to the modes of the multi-mode fiber by writing an analytic description of the propagation of the pulse in the system. For simplicity, we use the modes of a step-index multi-mode waveguide as our mathematical basis. We will limit our basis to the family of linearly polarized modes, under the weakly guided approximation (a full description of the modes can be found in [22]). Thus, we can construct a transfer operator which will evolve an incoming single mode field, in the form of a pulse emerging from the single-mode fiber, through the multi-mode fiber. We write it in the following form:

$$(1) \quad T\left(r,\theta,z,\omega\right) = \sum_{l,m} \alpha_{l,m}\left(\omega\right) E_{l,m}\left(r,\theta,z,\omega\right)$$

where $E_{l,m}$ are the spatial profiles of the $l,m$ mode including also the relevant phase accumulation along the propagation axis $z$, and $\alpha_{l,m}$ are complex valued coefficients representing the coupling created by the diffuser coupler and the single mode field input. These coefficients represent both magnitude and phase coupling. Therefore, from this point onwards we can disregard the spatial profile of the incoming field and let $\alpha_{l,m}$ encode this information directly.

Using this transfer operator, we can write the complete field evolving in the multi-mode fiber, given an arbitrary input field (of a given power spectrum and spectral phase):

$$(2) \quad \begin{aligned} E'\left(r,\theta,z,\omega\right) &= E_{input}\left(\omega\right) \cdot T\left(r,\theta,z,\omega\right) \\ &= A\left(\omega\right) e^{i\phi\left(\omega\right)} \cdot \sum_{l,m} \alpha_{l,m}\left(\omega\right) E_{l,m}\left(r,\theta,z,\omega\right) \end{aligned}$$

For simplicity, in our analysis and results, we will focus on obtaining the amplitude and phase of the pulse entering the multi-mode-fiber, noted as $E_{input}\left(\omega\right)$. As the propagation in the single-mode fiber is a linear operation, it can be readily back propagated to relate back to the original pulse and thus we can disregard this effect for our needs. Using (2) we write an analytic description of the sum frequency field generated in the nonlinear crystal, and the resultant interference pattern measured by camera:

$$(3) \quad M_{nonlinear}\left(x,y\right) = \int_{-\infty}^{\infty} I^2\left(x,y,t\right) dt = \int_{-\infty}^{\infty} \left| E\left(x,y,t\right) \right|^4 dt$$

We want to describe the nonlinear interference process described by Eq. (3) through the spectral fields. For a linear process this is straightforward by using Parseval's theorem, while for this nonlinear interference we take a direct approach and write the inverse Fourier transform:

$$(4) \quad \begin{aligned} M_{nonlinear}\left(x,y\right) &= \int_{-\infty}^{\infty} \left| \int_{-\infty}^{\infty} E\left(r,\theta,\omega\right) e^{i\omega t} d\omega \right|^4 dt \\ &= \int_{-\infty}^{\infty} \left| \int_{-\infty}^{\infty} A\left(\omega\right) e^{i\phi\left(\omega\right)} \cdot T\left(r,\theta,\omega\right) e^{i\omega t} d\omega \right|^4 dt \end{aligned}$$

This is the first indication that the nonlinearity of the sum frequency generation process creates a non-trivial functional dependence on the spectral phase. It is important to note that the functional that maps an input pulse to the nonlinear interference is not injective, which means that the inversion of this system contains ambiguities. This can be seen easily by exploring two common trivial functional ambiguities: a global phase and a time shift (or a linear phase in the Fourier domain). But these ambiguities are trivial and generally of no practical importance, hence their effect can be mitigated by working in a predefined subspace of pulses with equal global phase and a fixed time shift.

## 3. Reconstruction method

To the best of our knowledge, there is no analytic solution for the extraction of spectrum and spectral phase from the spatiotemporal nonlinear interference pattern. This poses the question on how can we reconstruct the incoming pulse out of the recorded intensity pattern. In principle, this is a regression problem and it can be stated as: given an interference pattern, we want to regress and find the pulse which created it. Mathematically, we can write the relation between a pulse and the interference pattern arising from that pulse, as the following functional:

$$(5) \quad \Gamma : E\left(\omega\right) \rightarrow M\left(r,\theta\right) \qquad \begin{aligned} E\left(\omega\right) &\in \mathbb{C}^n \\ M\left(r,\theta\right) &\in \mathbb{R}^{k \times k} \end{aligned}$$

where $M\left(r,\theta\right)$ is a matrix representing the discrete sampled interference pattern, $E\left(\omega\right)$ is a vector representing the discrete sampled complexed valued pulse, and $\Gamma$ is the integral derived in equation (4) In the regression problem we are trying to find the inverse of this relation.

We will take an approach similar to the one taken in [8] and construct a deep neural network to solve the regression problem to reconstruct the pulse from the recorded data. The deep neural network is a parametric function that can represent high-dimensional nonlinear functions. By learning the parameters of the network we are able to train the network to solve specific tasks. This concept has been around for some time now [23], but only in recent years, with the exponential growth in computational power and improved network architectures [24], deep networks are proving to be a formidable technique that can solve difficult and complex problems. By constructing an optimization problem on the parameters of a given network and using ground truth inputs and outputs (samples and labels), we can learn the correct parameters by solving the underlying optimization problem. Usually, direct solution is not possible due to the complexity of the problem, hence the problem is solved by iterations, using a variant of SGD (Stochastic Gradient Descent). In our problem, we pass an input image, i.e. the nonlinear interference pattern, through the computational layers of the network and receive an output vector which represents the temporal electric field.

This process of optimizing the network parameters is referred to as the training stage of the network. During this stage, it is important to avoid overfitting the parameters of the trained network to the specific data used for training, because tight overfitting might hamper the ability of high quality recovery of pulses other than those already included in the training set. Hence, to prevent overfitting, we train the network on data from the training set, and then test its performance on "test data" (which is not included in the training set and is used only to evaluate its performance) iteratively, until we find the best network parameters that yield the best fit on the testing data. In this way, we avoid overfitting and find the most suitable network parameters to recover the pulse shapes.

In our problem, we train the network to approximate the inversion of the functional described in Eq. (6) by feeding into the network nonlinear interference patterns as input and the pulses as labels, and we aim to minimize the L1 loss between the output of the network and the given label, i.e.:

$$w = \arg\min_{w'} \left\| DNN\left(I; w'\right) - E\left(t\right) \right\|_1 \qquad (6)$$

Where $w$ are the parameters we wish to optimize, $DNN\left(I; w'\right)$ is the output of the network, $I$ is the sum frequency generation interference pattern and $E\left(t\right)$ is the shape of the temporal pulse shape.

In practice, we will define our label as the real and imaginary part of the pulse, and concatenate them into a single vector. This means that we will try and find an $\mathbb{R}^{2n}$ vector rather than a $\mathbb{C}^n$ vector. This representation is less ambiguous than separating into amplitude and phase, and has shown better results in previous works [8].

Figure 2 shows the optimized architecture of our regression network used for reconstruction of pulses from nonlinear interference patterns. The network is composed of two major components, convolutional neural network (CNN) and a rectifier linear unit (ReLU). A CNN is a type of neural network which preforms cross-correlation between an input object and a learned kernel, meaning that each output value is composed of a local weighted average defined by the kernel. ReLU is a type of nonlinear activation function with the following form:

$$f\left(x\right) = \max\left(0, x\right) \qquad (7)$$

Our network is constructed by a 4-layer CNN followed by two fully connected layers with ReLU nonlinear activation between them. The figure also shows the dimension, channels and filter sizes in each layers.

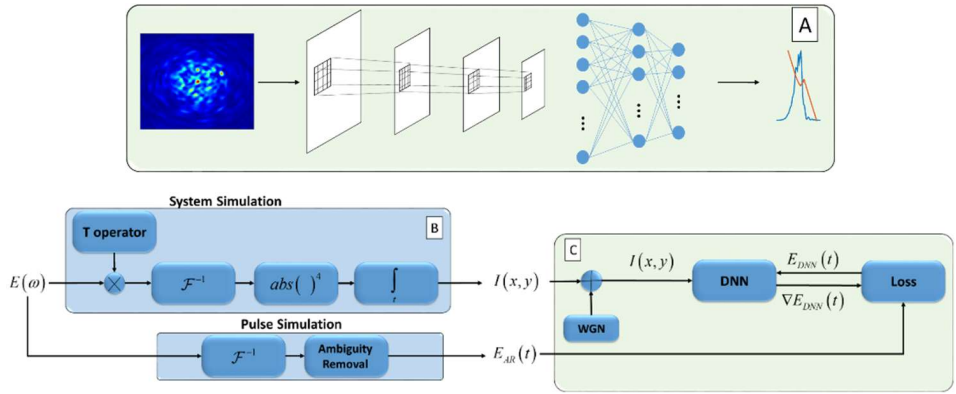

**Fig. 2** (A) Regression network architecture: four CNN layers followed by three fully connected layers. The input to this network is a sum frequency interference pattern, which is passed through the computational layers, until a final output is produced in the form of a vector of the real and imaginary parts of the temporal electric field. (B) Block diagram of sum frequency interference measurement and of the label generation from an input spectral pulse, as described in section 2 and section 3. The input of the simulation is passed on to the regression network. (C) Supervised training of the regression network. Each interference pattern is passed through the network to create a reconstruction. The error between a reconstructed

pulse and its ground truth pulse shape is used in back propagation and gradient descent to train the network and improve the network parameters.

## 4. Results

In order to learn the functional mapping described in equation (6) we need to create a dataset representing the physical relation between the pulse and the nonlinear interference pattern measured by the camera. Our simulated system consists of a multi-mode fiber of length 1cm, core radius of 50µm and refraction indices of $n_{core} = 1.4699, n_{clad} = 1.4533$ (parameters of a commercially available multimode fiber). Using these parameters we construct a transfer operator in the manner described in (1), with $l \in [0,19]$, $m \in [0,10]$ and we choose the coupling coefficients, $\alpha_{l,m}$, as $\alpha_{l,m} = e^{ix}$ where $x \sim Uni[0,2\pi]$. In a real physical system each coefficient will have a slightly different amplitude with some dependence on the wavelength; however, for a large enough number of modes in the multi-mode fiber, we assume that this choice of coupling coefficient is reasonable for a proof-of-concept demonstration.

In our simulations, we use pulses of a Gaussian power spectrum with a randomly generated spectral phase. The spectral phase is smoothed by removing high frequencies in order to ensure the pulse to be time limited yet complex, selected to have pulses with time bandwidth product in the range of $[0.5,5]$. As noted, to be able to invert the functional relation, we remove the trivial phase and time shift ambiguities by setting the phase at the center of the pulse to be zero and centering the max amplitude of the pulse to the center of the vector.

Working in the digital domain, we discretize the above functions using a linearly spaced spectral grid points in the range of $\omega \in [1.93Phz, 3.01Phz]$ with spacing of $\Delta\omega = 4.24Thz$, equivalent to temporal resolution of $5.8fs$, corresponding to 256 spectral points and wavelength span in the range $\lambda \in [625nm, 975nm]$. The 2D linear spatial grid is chosen such that each pixel is of the size of $0.21\mu m \times 0.21\mu m$ with $256 \times 256$ spatial points corresponding to $(x,y) \in [-27.5\mu m, 27.5\mu m] \times [-27.5\mu m, 27.5\mu m]$, preserving resolution in regions of interest yet keeping computational effort low.

To train the system, we create a dataset of 100,000 pulses and their corresponding nonlinear interference pattern, as described in section 2. Through random sampling of pulses from the train set over many iterations, we pass the interference patterns through the network. This gives us the reconstructed pulses. By computing the error between reconstruction and ground truth pulses, we back-propagate the error and optimize the network parameters, by the process of gradient descent. In practice, we use a popular variant of gradient descent named ADAM [25]. This process is illustrated in figure 2. By this process of training, we create two models (the set of parameters that describe the network): one trained without noise present in measurements and one trained with an additive white Gaussian noise of SNR $\sim Uni[0dB, 30dB]$.

Having trained the network, we now test the trained models performance by creating a test set of 6500 pulses, which are not part of the training set, that is, the test pulses have not been previously seen by the network. By passing the interferences images of the test set pulses through the net, we obtain a set of reconstructed pulses. Examples of such test pulses and their reconstructions can be seen in figure 3 for the noiseless model and with no noise present in the test set measurements. Using the L1 error criterion between ground truth pulses and reconstructed pulses on noiseless measurements, on average we obtain a final reconstruction error of $0.83 \times 10^{-3} \pm 0.69 \times 10^{-3}$. This error is very low, compared to the pulses norm which is normalized to unity. We also compare the reconstruction quality in terms of the angle, defined as $\delta_\theta = \cos^{-1}\left(\frac{\langle E_G, E_R \rangle}{\|E_G\| \cdot \|E_R\|}\right)$ between the ground truth and the reconstructed pulses, in similar fashion to [8]. We obtain $\delta_\theta = 0.074 \pm 0.062$ which indicates good reconstruction quality as on average each reconstructed pulse is highly correlated with its corresponding ground truth.

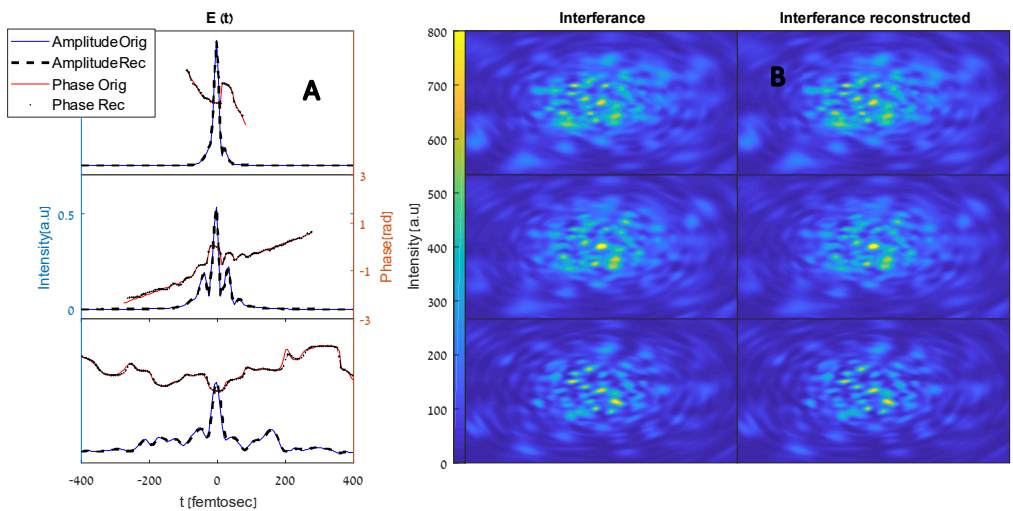

**Fig. 3** Three examples of pulse reconstruction with varying complexity: Top simple, middle medium, bottom most complex. (A) Temporal amplitude and phase of an original and reconstructed pulse. (B) The nonlinear interference of the original pulse and its reconstructed counterpart.

We also compare the trained models performance with and without AWG noise present in test set measurements, which can be seen in figure 4. As expected, when SNR is high the two models achieve similar accuracy, yet once noise is introduced and the SNR is lowered - the model trained with noise shows greater noise immunity and reconstruction accuracy, to the extent of 30% improvement.

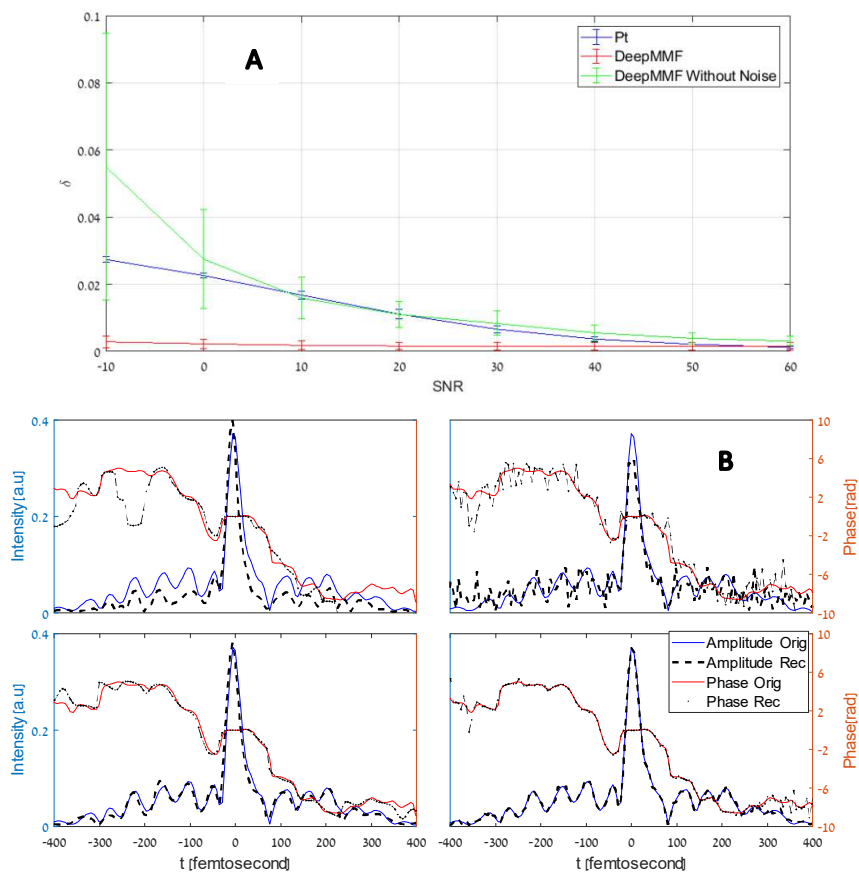

**Fig. 4** (A) Error as a function of SNR for reconstruction of 30 pulses using FROG ptychography reconstruction, and using the two DeepMMF models: one trained with AWG noise and the other without noise. The bars are $\sigma_{STD}$ of error. (B)

Reconstruction examples for one pulse, using the two reconstruction algorithms and at two SNR points. Left Colum displays DeepMMF reconstruction, right is ptychography. Top Row is at 10dB SNR, Bottom 40dB SNR.

Next, we test the robustness to transfer operator estimation mismatch, i.e. the sensitivity of the network to errors in the operator. This test is motivated by the algorithmic challenge this system introduces: the transfer coefficient of a real lab coupler is unknown, hence, in order to simulate a real system, one would need to measure or estimate these coefficients. This will be further elaborated in the discussion. For the sake of comparison, we create 3 instances of operators: a ground truth operator, an operator whose $\alpha_{l,m}$ coefficients differ by 1% and an operator with 10% difference from the ground truth. The difference is generated by adding a $x\%$ uniform random phase to the coefficients. We train a network using the train set from the ground truth operator and test it using test sets from the three operators. The results of this test can be seen in figure 5. We can see that the 1% difference in coefficients created a small shift in the histogram towards higher errors which means there was a decrease in performance. As this degradation is small, compared to variance of the errors within the pulse distribution, it shows that the system is robust to small deviations in the physical elements.

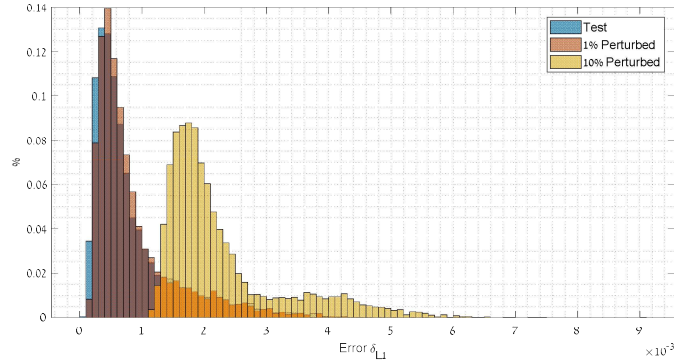

**Fig. 5** Probability distribution of L1 reconstruction errors of the three test sets, done on 6500 pulses.

## 5. Conclusions

We proposed a new simple all in line system for ultrashort pulse reconstruction from sum frequency field interference measurements, employing deep learning to numerically invert the nonlinear interference pattern to pulse mapping, as a method for analytical inversion is unknown. Using simulated data, we have shown that this method achieves good reconstruction results, with an average error of $0.02 \pm 0.01\%$.

The system shows good performance even in scenarios of low SNR, which is a key property for single-shot ultrashort pulse measurement system. We expect our method to be especially useful in cases where single-shot measurements are required and there is a power constraint on the pulses employed (e.g., when studying biological specimen).

As the transfer operator of the presented device is system-dependent, the training of the network must be done on a **specific** real system. This can be achieved by either taking interference measurements of known pulses (labels) from the system itself or by simulation of measurements of simulated pulses using the parameters of the system. Naturally from the system itself or by simulation of measurements. Naturally, using simulated data (instead of data from experiments) for training the network is highly desirable, but this may pose challenges due to discrepancies between real and simulated data (problem known as "sim to real" in deep networks). Therefore, showing that the system is robust to small errors in the transfer operator implies that we can rely on estimating the parameters of the fiber and the coupler, rather than measuring them precisely, and thereby train the network on simulated pulses.

Before closing, we note that the system proposed here introduces a new feature with respect to making use of the redundancy of the measurements. Namely, the recoded data in FROG and d-scan is generally redundant. That is, there are many more measurements than required for reconstructing the pulse uniquely. This redundancy makes these techniques powerful, giving rise to robustness to noise and misalignment as well as allowing reconstructing a pulse at higher temporal resolution from the temporal scanning steps in FROG [17] and reconstructing multiple pulses from a single FROG trace [18,19]. However, the measured data for each scanning step in FROG or for specific dispersion in d-scan is not redundant; in fact, it does not even contain enough information to reconstruct the pulse. Here comes into play another added value of our technique: in our system, due to the strong spatiotemporal mixing in the MMF, the temporal information of the pulse is mixed completely in the 2D image. There is no separation into rows or columns. Hence, in principle, every subset of the image is highly redundant. This endows our technique with possibly greater robustness that can make it advantageous for real time probing of ultrafast processes and under noisy conditions. Also, it is important to note that our technique does not suffer from the stringent physical restrictions characteristic of other single-shot methods, for example, on the time-bandwidth product [14] or on the spatial beam profile [20], avoiding these limitations facilitates greater dynamic range and robustness.

Finally, it is worth noting that the architecture of the network can be further improved by adding a generative model, such as Generative Adversarial Network (GAN) [26], which can be used for enriching the distribution of pulses learned and thus improving robustness of reconstruction. Thus, we believe that this method can achieve state of the art results, comparable to FROG reconstruction accuracy, while improving the noise immunity and thus improving SNR.

## References

1. S. X. Hu and L. A. Collins, "Attosecond Pump Probe: Exploring Ultrafast Electron Motion inside an Atom," Phys. Rev. Lett. **96**, 073004 (2006).
2. W. Demtroder, "Laser Spectroscopy: Basic Concepts and Instrumentation, Second Enlarged Edition," Opt. Eng. **35**, 3361 (1996).
3. K. E. Sheetz and J. Squier, "Ultrafast optics: Imaging and manipulating biological systems," J. Appl. Phys. **105**, 051101 (2009).
4. R. Trebino, *Frequency-Resolved Optical Gating: The Measurement of Ultrashort Laser Pulses* (Springer US, 2000), Vol. 62.
5. M. Miranda, C. L. Arnold, T. Fordell, F. Silva, B. Alonso, R. Weigand, A. L'Huillier, and H. Crespo, "Characterization of broadband few-cycle laser pulses with the d-scan technique," Opt. Exp. **20**, 18732 (2012).
6. A. Krizhevsky, I. Sutskever, and G. E. Hinton, "ImageNet Classification with Deep Convolutional Neural Networks," Adv. Neural Inf. Process. Syst. 1–9 (2012).
7. A. Esteva, B. Kuprel, R. A. Novoa, J. Ko, S. M. Swetter, H. M. Blau, and S. Thrun, "Dermatologist-level classification of skin cancer with deep neural networks," Nature **542**, 115–118 (2017).
8. T. Zahavy, A. Dikopoltsev, D. Moss, G. I. Haham, O. Cohen, S. Mannor, and M. Segev, "Deep learning reconstruction of ultrashort pulses," Optica **5**, 666–673 (2018).
9. S. Kleinert, A. Tajalli, T. Nagy, and U. Morgner, "Rapid phase retrieval of ultrashort pulses from dispersion scan traces using deep neural networks," Opt. Lett. **44** 979 (2019).
10. J. White and Z. Chang, "Attosecond streaking phase retrieval with neural network," Opt. Exp. **27** 4799 (2019).
11. P. Vincent and H. Larochelle, "Stacked Denoising Autoencoders: Learning Useful Representations in a Deep Network with a Local Denoising Criterion Pierre-Antoine Manzagol," J. Mach. Learn. Res. **11**, 3371–3408 (2010).
12. C. Horn, M. Wollenhaupt, M. Krug, T. Baumert, R. De Nalda, and L. Bañares, "Adaptive control of molecular alignment," Phys. Rev. A **73** 031401 (2006).
13. Y.-H. Chen, S. Varma, I. Alexeev, and H. Milchberg, "Measurement of transient nonlinear refractive index in gases using xenon supercontinuum single-shot spectral interferometry," Opt. Exp. **15** 7458 (2007).
14. P. O'Shea, M. Kimmel, X. Gu, and R. Trebino, "Highly simplified device for ultrashort-pulse measurement," Opt. Lett. **26**, 932 (2001).
15. D. Fabris, W. Holgado, F. Silva, T. Witting, J. W. G. Tisch, and H. Crespo, "Single-shot implementation of dispersion-scan for the characterization of ultrashort laser pulses," Opt. Express **23**, 32803 (2015).
16. C. Iaconis and I. A. Walmsley, "Spectral phase interferometry for direct electric-field reconstruction of ultrashort optical pulses," Opt. Lett. **23**, 792 (1998).
17. P. Sidorenko, O. Lahav, Z. Avnat, and O. Cohen, "Ptychographic reconstruction algorithm for frequency-resolved optical gating: super-resolution and supreme robustness," Optica **3**, 1320 (2016).
18. C. Bourassin-Bouchet and M. E. Couprie, "Partially coherent ultrafast spectrography," Nat. Commun. (2015).
19. G. I. Haham, P. Sidorenko, O. Lahav, and O. Cohen, "Multiplexed FROG," Opt. Exp. **25**, 33007 (2017).
20. D. Fabris, W. Holgado, F. Silva, T. Witting, J. W. G. Tisch, and H. Crespo, "Single-shot implementation of dispersion-scan for the characterization of ultrashort laser pulses," Opt. Exp. **23** 32803 (2015).
21. B. Redding and H. Cao, "Using a multimode fiber as a high-resolution, low-loss spectrometer," Opt. Lett. **37**, 3384–3386 (2012).
22. K. Okamoto, *Fundamentals of Optical Waveguides* (2006).
23. D. E. Rumelhart, G. E. Hinton, and R. J. Williams, "Learning representations by back-propagating errors," Nature **323**, 533–536 (1986).
24. Y. LeCun, K. Kavukcuoglu, and C. Farabet, "Convolutional networks and applications in vision," in *ISCAS 2010 - 2010 IEEE International Symposium on Circuits and Systems: Nano-Bio Circuit Fabrics and Systems* (2010), pp. 253–256.
25. D. P. Kingma and J. L. Ba, "Adam: A method for stochastic gradient descent," ICLR Int. Conf. Learn. Represent. 1–15 (2015).
26. I. Goodfellow, J. Pouget-Abadie, M. Mirza, B. Xu, D. Warde-Farley, S. Ozair, A. Courville, and Y. Bengio, "Generative Adversarial Nets," Adv. Neural Inf. Process. Syst. 27 2672–2680 (2014).